\documentclass{INTERSPEECH2023}
\usepackage{graphicx}
\usepackage{bm}
\usepackage{helvet}
\usepackage{courier}
\usepackage{amsfonts,amssymb}
\usepackage{amsmath,amsfonts,amsthm}
\usepackage{booktabs}
\usepackage{url}
\usepackage{soul}
\usepackage{color}
\usepackage{flushend}

\newcommand{\etal}{\textit{et al.}}
\interspeechcameraready 

\usepackage{balance}

\title{Dual Transformer Decoder based Features Fusion Network \\ for Automated Audio Captioning} 

\name{Jianyuan Sun$^1$, Xubo Liu$^1$, Xinhao Mei$^1$, Volkan Kılıç$^2$, Mark D. Plumbley$^1$, Wenwu Wang$^1$}

\address{
  $^1$Centre for Vision, Speech and Signal Processing (CVSSP), University of Surrey, UK \\
  $^2$Department of Electrical and Electronics Engineering, Izmir Katip Celebi University, Turkey}
\email{\{jianyuan.sun, xubo.liu, x.mei, m.plumbley, w.wang\}@surrey.ac.uk, vkilic2004@gmail.com}

\begin{document}

\maketitle
 
\begin{abstract}
Automated audio captioning (AAC) which generates textual descriptions of audio content. Existing AAC models achieve good results but only use the high-dimensional representation of the encoder. There is always insufficient information learning of high-dimensional methods owing to high-dimensional representations having a large amount of information. In this paper, a new encoder-decoder model called the Low- and High-Dimensional Feature Fusion (LHDFF) is proposed. LHDFF uses a new PANNs encoder called Residual PANNs (RPANNs) to fuse low- and high-dimensional features. Low-dimensional features contain limited information about specific audio scenes. The fusion of low- and high-dimensional features can improve model performance by repeatedly emphasizing specific audio scene information. To fully exploit the fused features, LHDFF uses a dual transformer decoder structure to generate captions in parallel. Experimental results show that LHDFF outperforms existing audio captioning models.
\end{abstract}


\section{Introduction}
Automated audio captioning (AAC) is a cross-modal translation task that generates a text description for a given audio clip~\cite{DrossosAV17, mei2022ac_review}. The problem of AAC has recently received much attention in the fields of acoustic signal processing and machine learning due to its potential applications such as describing audio content for hearing impaired people and generating text descriptions for audio search, retrieval, and indexing~\cite{mei2022language,liu2022simple, lass, mei2022metric, liu2021conditional}. 

Conventional AAC models are generally based on an encoder-decoder architecture in which an encoder is used to extract the latent embedding from the audio clip and a decoder is used for text generation based on the audio embedding~\cite{DrossosAV17, koizumi2020transformer, MeiHLCWWZLKTSPW21}. In the architectures used for the encoder, pre-processing is often employed to extract useful information for the latent embedding representation. Tran \etal~proposed an encoder with three learnable processes to extract and fuse local and temporal information~\cite{tran2020wavetransformer}. Xu \etal~explored a transfer learning method to learn local and global information~\cite{xu2021investigating} in which audio tagging (AT) and acoustic scene classification (ASC) were employed to represent local and global audio information. Mei \etal~employed a transformer encoder to process audio information, which can capture the temporal relationship among audio events~\cite{MeiLHPW21}. Moreover, Liu \etal~employed a contrastive loss to cluster the representations of the audio-text paired data in latent space while distinguishing between unpaired negative data~\cite{LiuHMKTPW21}. After that, Chen \etal~proposed an interactive audio-text representation method for the audio encoder using contrastive learning~\cite{chen2022interactive}. With the popularity of multimodal approaches, Liu \etal~investigated to the introduction of visual modality for improving the performance of existing audio captioning systems~\cite{liu2022visually}. Mei \etal \cite{mei2021diverse, mei2022towards} proposed to utilize the adversarial learning method to generate diverse audio captions for an audio clip. The majority of existing methods mentioned above only use high-dimensional representation output by the encoder as the input of the decoder to generate the captions. 

AAC is a highly challenging task, as existing methods may not be able to fully learn the vast amount of information present in high-dimensional representations. Using low-dimensional representations can address the issue of existing methods being unable to fully learn the vast amount of information in high-dimensional representations. However, low-dimensional features only contain limited information for a fixed number of specific audio scenes. To better balance the problem of low- and high-dimensional features, in this paper, a new encoder-decoder model called the Low- and High-Dimensional Feature Fusion (LHDFF) is proposed. In LHDFF, a new encoder is used called Residual PANNs (RPANNs) by fusing the low- and high-dimensional features generated in the intermediate and final convolution blocks, respectively. This fusion allows the low- and high-dimensional features to cooperate and complement each other, which can emphasize the recurring audio scenes (as observed in Section~\ref{sec:resultss}). We then obtain the final captioning by using a fused probabilistic approach, leading to more accurate captions due to the complementary nature of low- and high-dimensional features.

The remainder of the paper is organized as follows. Our proposed LHDFF model is introduced in Section~\ref{sec:proposed method}. Section~\ref{sec:experiments} shows experimental results. Finally, Section~\ref{sec:conc} draws conclusions.

\section{Proposed method}
\label{sec:proposed method}
\begin{figure}[!hbt]
  \centering
  \includegraphics[height = 7.73cm, width = 7.85cm]{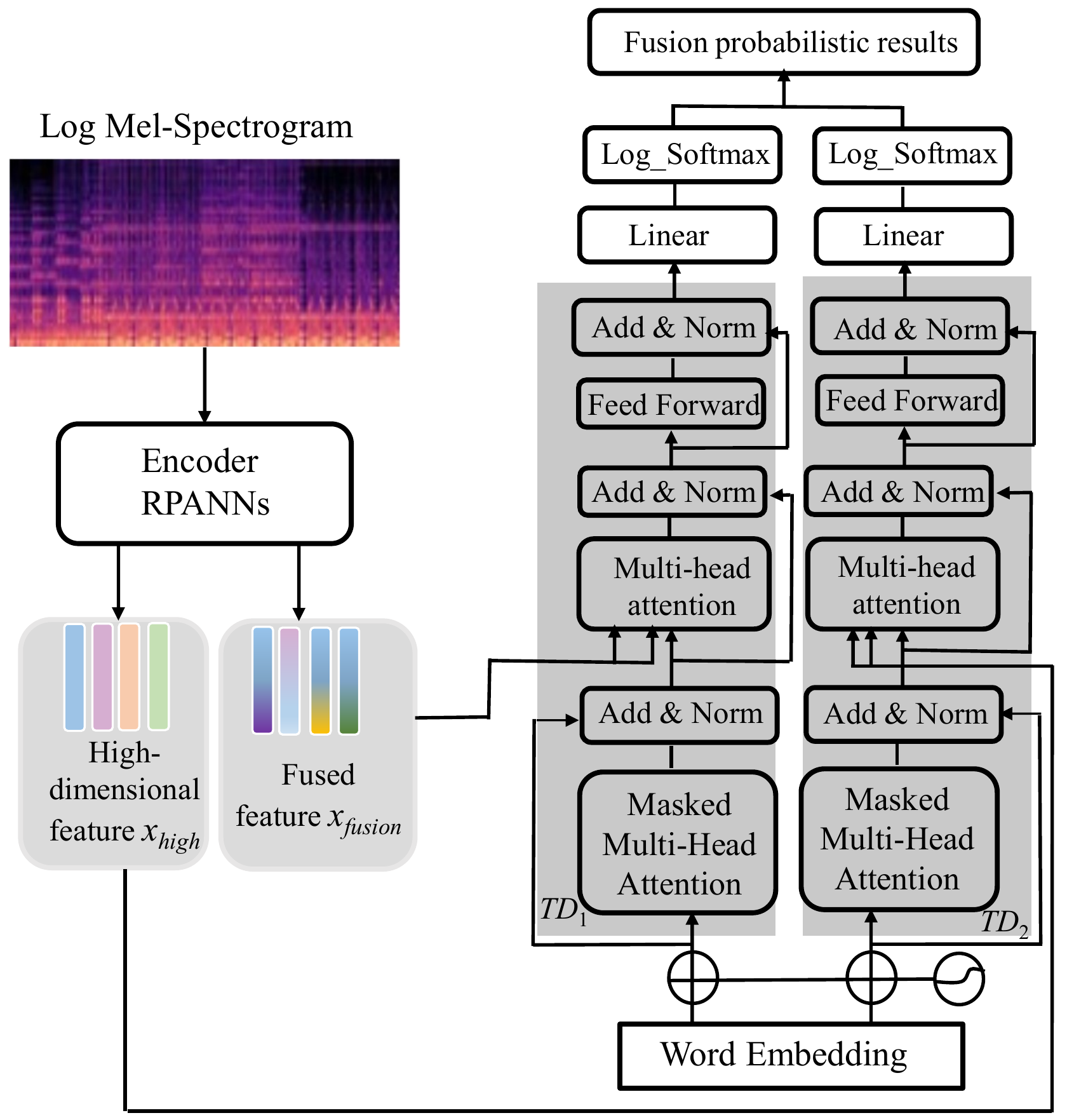}
  \vspace{1em}
  \caption{The architecture of the proposed LHDFF. LHDFF consists of an RPANNs encoder and a dual decoder. The RPANNs encoder outputs two embedding features, i.e., the high-dimensional feature $x_{high}$ and the fusion feature $x_{fusion}$. Moreover, $x_{fusion}$ and $x_{high}$ are input into the dual transformer decoder $TD_{1}$ and $TD_{2}$, respectively. The dual decoder consists of two transformer decoders, i.e., $TD_{1}$ and $TD_{2}$. Each transformer decoder $TD_{1}$ and $TD_{2}$ follows a linear layer and a log softmax layer to output the probability distribution along the vocabulary vector. The final fused probability result is obtained by fusing the log softmax probability distribution of the transformer decoder $TD_{1}$ and $TD_{2}$.}\label{fig:MODEL}
\end{figure}

In this section, we present the proposed LHDFF architecture that consists of the RPANNs encoder and the dual transformer decoder. The overall LHDFF architecture is depicted in Fig.~\ref{fig:MODEL}, where, in the encoder part, the low- and high-dimensional features are fused to cooperate and complement each other.

\subsection{RPANNs encoder}

Classical PANNs have used the CNN10 model, which consists of four convolutional blocks, and each block has two convolutional layers with a $3\times3$ kernel size, followed by batch normalization and ReLU. The channel number of each convolutional block is $64, 128, 256$, and $512$. Moreover, an average pooling layer with the kernel size $2\times2$ is applied for downsampling. After the last convolutional block, a global average pooling is applied along the frequency axis to align the dimension of the output with the hidden dimension $D$ of the decoder.  

\begin{figure}[!hbt]
  \centering
  \includegraphics[height = 7.83cm, width = 4.45cm]{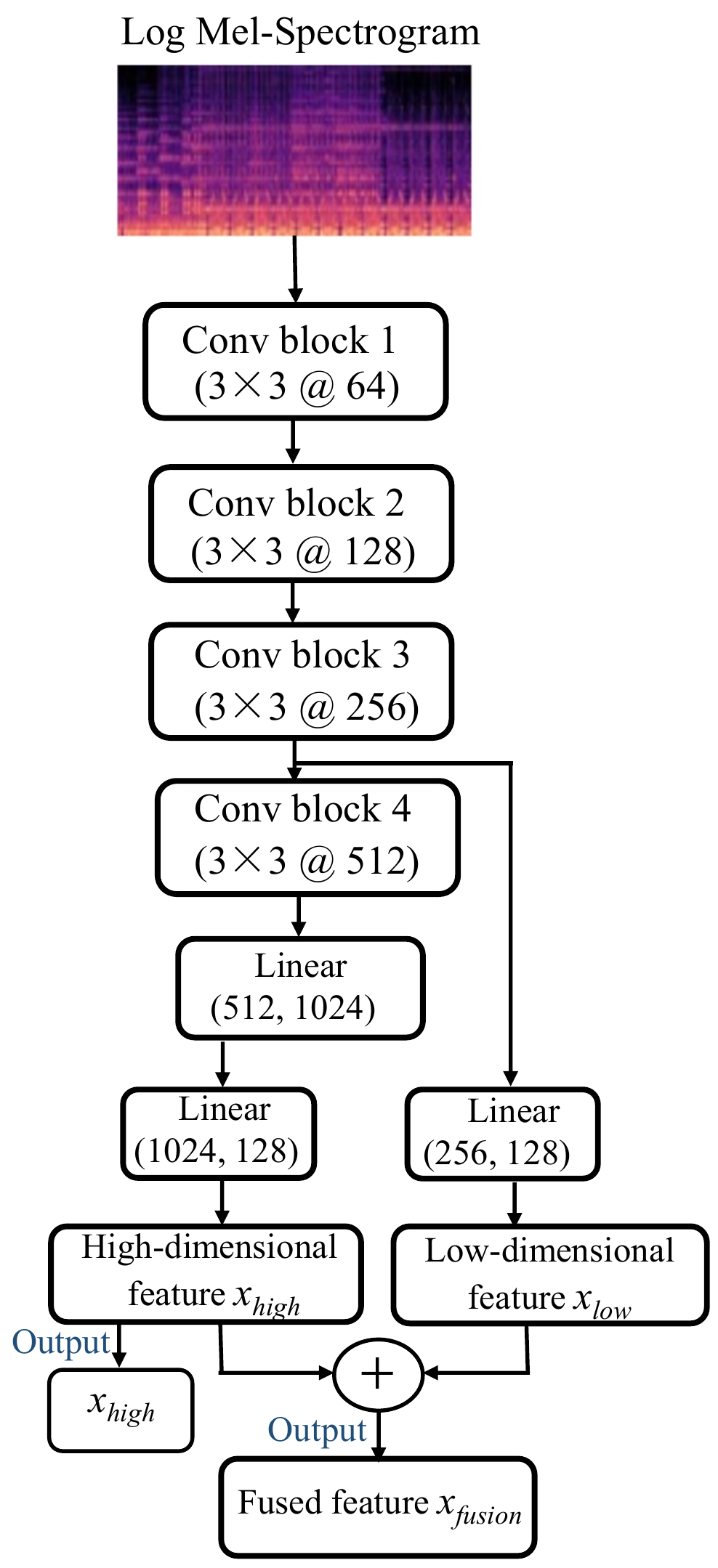}
   \caption{The architecture of the proposed RPANNs. RPANNs consist of four convolutional blocks. The RPANNs encoder outputs two features, i.e., the high dimensional feature $x_{high}$ and the fused feature $x_{fusion}$. Where the fused feature $x_{fusion}$ is fused by the low-dimensional feature from the third convolutional block and the high-dimensional feature $x_{high}$.}\label{fig:RPANNS}
\end{figure}

To further improve the performance of PANNs, we propose a new PANNs encoder called RPANNs. The basic architecture of the proposed RPANNs is similar to PANNs. In Fig.~\ref{fig:RPANNS}, we can see the RPANNs consist of four convolutional blocks. The channel number $D$ of each convolutional block is $64, 128, 256$, and $512$. The difference between PANNs and RPANNs is that RPANNs fuse the low-dimensional feature from the third convolutional block and the high-dimensional feature from the final layer. The RPANNs encoder takes the log mel-spectrogram of an audio clip as the input and generates the high dimensional feature $I\in \mathbb{R}^{T\times D}$, where $T$ denotes the numbers of time frames, and $D$ represents the dimension of the spectral features at each time frame. Let $x_{3}$ be the output of the third convolutional block indicating a low dimensional feature. Then, $x_{3}\in\mathbb{R}^{T^{'}\times D}$, where $D = 256$, i.e. $x_{3}\in\mathbb{R}^{T^{'}\times 256}$. Let $x_{final}$ denote the output of the final layer, which is a high-dimensional feature. The final layer is a linear layer with $1024$ dimensional, then the high-dimensional feature $x_{final}\in\mathbb{R}^{T\times D}$, where $D = 1024$, i.e. $x_{final}\in\mathbb{R}^{T\times 1024}$.

In RPANNs, we set the dimension of the high-dimensional feature $D = 128$. The low- and high-dimensional fused feature $x_{fusion}$ and high-dimensional feature $x_{high}$ are summarized as follows:
\begin{equation}
\begin{split}
x_{high} &= Relu(f_{128}(x_{final})), \qquad x_{high}\in \mathbb{R}^{T\times 128}; \\
x_{low} & = Relu(f_{128}(x_{3})), \qquad  \qquad x_{low}\in \mathbb{R}^{T^{'}\times 128};  \\
x_{fusion} &= x_{high} \oplus x_{low}, \qquad \qquad x_{fusion}\in \mathbb{R}^{T\times 128}.
\end{split}
\end{equation}
As $T^{'}\ne T$, the technology of filling $0$ is used to make the dimension of $T^{'}$ and $T$ the same. After that, $x_{fusion}$ and $x_{high}$ are fed into the dual decoder at the same time. 

\subsection{Dual decoder}

We design a dual decoder to exploit the information of the fused feature, $x_{fusion}$, and high-dimensional feature $x_{high}$. The dual decoder is based on two standard transformer decoders consisting of eight parts, a word embedding layer, two standard Transformer decoders, two linear layers, two softmax probability layers, and one fusion layer, as shown in Fig.~\ref{fig:MODEL}. The word embedding layer is represented with $V \times d$ word embedding matrix $W$, where $V$ denotes the size of the vocabulary and $d$ is the dimension of each word vector. The word embedding layer is frozen in training after random initialization.

The two transformer decoders are denoted as $TD_{1}$ and $TD_{2}$, with $W$, together with $x_{fusion}$ and $x_{high}$, as their inputs, respectively. Each transformer decoder includes a multi-head attention mechanism. Empirically, we set the number of attention heads as $4$ for each transformer decoder, and the dimension of the hidden layer is $128$. Both transformer decoders have a multi-head attention mechanism in which the number of heads and the size of the hidden layer is set to four and $128$, respectively, based on extensive empirical evaluations. In addition, $TD_{1}$ and $TD_{2}$ are followed by linear layers $f_{TD_{1}}$ and $f_{TD_{2}}$, and a softmax probability layer to output the probability distribution along the vocabulary vector.

Let $m$ represent the number of words in the dictionary for a dataset. Then, the outputs of linear layers $x_{TD_{1}}$ and $x_{TD_{2}}$ are calculated as follows:
\begin{equation}
\begin{split}
x_{TD_{1}} &= f_{TD_{1}}(x_{fusion}), \qquad x_{TD_{1}}\in \mathbb{R}^{T \times m}; \\
x_{TD_{2}} &= f_{TD_{2}}(x_{high}), \qquad x_{TD_{2}}\in \mathbb{R}^{T \times m}.
\end{split}
\end{equation}

The final fused probability distribution $P_{fusion}$ for the vocabulary can be calculated by fusing the probability distribution of each transformer decoder $P_{TD_{1}}$ and $P_{TD_{2}}$, as follows: 
\begin{equation}
\begin{split}
& P_{TD_{1}} = log(softmax(x_{TD_{1}})), \qquad P_{TD_{1}}\in \mathbb{R}^{T\times m};\\
& P_{TD_{2}} = log(softmax(x_{TD_{2}})), \qquad P_{TD_{2}}\in \mathbb{R}^{T\times m};\\
& P_{fusion} = P_{TD_{1}} \oplus P_{TD_{2}}, \qquad \qquad P_{fusion}\in \mathbb{R}^{T\times m}.
\end{split}
\end{equation}
Where the operator symbol $\oplus$ means that the corresponding elements in two probability distribution vectors are added. 

The training objective of RPANNs is to optimize the cross-entropy (CE) loss defined in terms of all the possible model parameters $\theta$ as:
\begin{equation}
L_{CE}(\theta) = -\frac{1}{T}\Sigma_{t=1}^{T} logP_{fusion}(y_{t}|y_{1:t-1}, \theta)
\end{equation}
where $y_{t}$ represents the ground truth word at time step $t$.

\section{Experiments}
\label{sec:experiments}

This section discusses experimental evaluations of the proposed LHDFF on Clotho \cite{drossos2020clotho} and AudioCaps \cite{kim2019audiocaps} datasets as compared with state-of-the-art baseline approaches. We start with a description of the datasets, pre-processing, experimental setups, and performance metrics, before giving the analysis and comparison of the results.

\subsection{Datasets}
Clotho~\cite{drossos2020clotho} is a well-known audio captioning dataset with audio clips collected from the Freesound archive. The audio clips are between $15$ and $30$ seconds long and contain five captions with $8$ to $20$ words annotated by different Amazon Mechanical Turk employees. In our experiment, the Clotho v2 is used that was released for Task $6$ of DCASE $2021$ Challenge, which contains $3839$ development, $1045$ validation, and $1045$ evaluation samples. To comply with the settings of the baseline methods, we also merge the development and validation samples, which refer to the training dataset with $4884$ audio clips. The evaluation split is selected as the test set with $1045$ audio clips. Moreover, each audio clip is combined one of its five captions as a training sample in the training set. 
\begin{table*}[tbp]
   \centering
  \caption{The comparative experimental results on Clotho and AudioCaps datasets. The Baseline method does not use the reinforcement learning (RL) strategy for fine-tuning.}
  \scalebox{0.9}{
  \begin{tabular}[]{c| c| c c c c c c c c c}
  \hline
  \textbf{Dataset}     &\textbf{Model}  &BLEU$_1$  &BLEU$_2$  &BLEU$_3$  &BLEU$_4$  &ROUGE$_L$  &METEOR  &CIDE$_r$ &SPICE  &SPIDE$_r$   \\ \hline
                        & Baseline~\cite{MeiHLCWWZLKTSPW21} (without RL) &0.561  &0.364  &0.243 &\textbf{0.159}    &0.375    &0.172  &0.391  &0.120    &0.256  \\
       Clotho           & CL4AC model~\cite{LiuHMKTPW21}   &0.553   &0.349    &0.226    &0.143    &0.374    &0.168   &0.368    &0.115    &0.242   \\
                        & AT-CNN10~\cite{xu2021investigating} &0.556   &0.363    &0.242    &0.159    &0.368    &0.169   &0.377    &0.115    &0.246   \\  \hline
                        &LHDFF~(nonfusion) &0.555 &0.363  &0.245  &0.161  &0.375  &0.173  &0.381  &0.117  &0.249 \\
                        &LHDFF~(fusion) &0.570 &0.370   &0.246  &0.158  &0.378  &0.174  &0.401  &0.120  &0.261 \\
                        &LHDFF~(fusion block2) & 0.565 &0.366 &0.245 &0.159 &0.377 &0.174 &0.392 &0.120 &0.256 \\
                        & LHDFF~ &\textbf{0.570} &\textbf{0.370} &\textbf{0.247} &\textbf{0.159}  &\textbf{0.378}  &\textbf{0.175} &\textbf{0.408} &\textbf{0.122} &\textbf{0.265} \\ \hline
                        & Baseline~\cite{MeiHLCWWZLKTSPW21}~(without RL) &0.667     &0.491     &0.350   &0.248    &0.468    &0.229  &0.643  &0.165 &0.404 \\ 
      AudioCaps         & Pre-Bert~\cite{Liuprebert22}  &0.667    &0.491    &0.354 & 0.247    & 0.475   &\textbf{0.232}   & 0.654    & 0.167 & 0.410 \\
                        & AT-CNN10~\cite{xu2021investigating} &0.655   &0.476   &0.335    &0.231    &0.467  &0.229  &0.660  &0.168  &0.414  \\ \hline
                        &LHDFF~(nonfusion) & 0.662 &0.484 &0.344  &0.244  &0.467 &0.226 &0.639 &0.167 &0.403 \\
                        & LHDFF~(fusion)  &\textbf{0.674}     &0.500     &0.367   &0.263  &0.481 &0.231  &0.666 &\textbf{0.171}    &0.419 \\  
                        &LHDFF~(fusion block2) & 0.668 &0.494 &0.358  &0.254  &0.479 &0.231 &0.663 &0.170 &0.417 \\ 
                        & LHDFF~ &\textbf{0.674} &\textbf{0.502} &\textbf{0.368} &\textbf{0.267} &\textbf{0.483} &\textbf{0.232} &\textbf{0.680} &\textbf{0.171} &\textbf{0.426} \\
                        \hline   
  \end{tabular}}
  \label{tab:audioresult}
\end{table*}

\begin{table}[tbp]
  \caption{The generated captioning results on a test audio clip of the AudioCaps dataset.}
  \begin{tabular}[]{c| c }
  \hline
  \textbf{Audio clip}     & Yti66RjZWTp0.wav   \\ \hline
Ground truth     & a \textbf{male speaks} as metal \textbf{clicks} and a \\ & \textbf{gun} fires once. \\ \hline       
Baseline                  & a man speaks followed by \\ & several gunshots. \\ \hline
LHDFF(our model)         & a \textbf{man speaks} followed by several \\ & loud                                             \textbf{clicks} and a \textbf{gun} shots. \\ \hline 
  \end{tabular}
  \label{tab:audio1}
\end{table}

AudioCaps~\cite{kim2019audiocaps} is the largest audio captioning dataset, which includes $50k$ audio clips with a duration of $10$ seconds. AudioCaps is divided into three parts with $49274$ audio clips for training, $497$, and $957$ audio clips for validation and testing, respectively. Each audio clip has one caption in the training set, and each audio clip involves $5$ captions in the validation and test sets, respectively. The length of the captions ranges from $3$ to $20$ words. 

\subsection{Data pre-processing}

For the audio clips, we use a 1024-point Hanning window with a hop size of 512 points to obtain 64-dimensional log mel-spectrograms as the LHDFF input features. Moreover, the SpecAugment~\cite{park2019specaugment} method is employed to augment the log mel-spectrogram of an audio clip in the training data using the “zero-value masking” and “mini-batch based mixture masking”~\cite{wang2021specaugment++}. The captions in the Clotho and AduioCaps datasets are converted to lowercase with punctuation removed. Moreover, we pad two special tokens “\textless sos\textgreater” and “\textless eos\textgreater” at the beginning and end of each caption.   

\subsection{Experimental setups}

The proposed LHDFF model is trained using the Adam optimizer~\cite{kingma2014adam} with a batch size of $32$. We set the model training epoch to $30$ with an initial learning rate (LR) of $5 \times 10^{-4}$, because the model performs best at epoch 30 on the validation set. In the first $5$ epochs, warm-up is applied to increase the initial LR linearly. Then, the LR is decreased to 1/10 every 10 epochs. For all the captioning, the Word2Vec model is used to pre-train the word embedding in Clotho and AudioCaps~\cite{mikolov2013efficient}.

\subsection{Performance metrics}

To evaluate the performance of AAC models, several metrics are employed including machine translation metrics: BLEUn~\cite{papineni2002bleu}, METEOR~\cite{agarwal2007meteor}, ROUGEL~\cite{rouge2004package} and captioning metrics: CIDEr~\cite{vedantam2015cider}, SPICE~\cite{anderson2016spice}, SPIDEr~\cite{liu2017improved}. BLEUn mainly measures the n-gram precision of a generated text while METEOR is a word-to-word matching-based harmonic mean of recall and precision. ROUGEL computes F-measures based on the longest common sub-sequence. The term frequency-inverse document frequency of the n-gram is used in the calculation of CIDEr. SPICE takes captions from scene graphs and uses them to determine F-score. SPIDEr is the mean of CIDEr and SPICE scores.

\subsection{Results}
\label{sec:resultss}

The proposed LHDFF model was evaluated in the Clotho dataset and compared with three representative baseline methods including conventional encoder-decoder based audio captioning system \cite{MeiHLCWWZLKTSPW21}, CL4AC model \cite{LiuHMKTPW21}, and AT-CNN10 \cite{xu2021investigating}. The baseline model~\cite{MeiHLCWWZLKTSPW21} contains a PANNs encoder and a transformer decoder, which uses reinforcement learning techniques.

For a fair comparison, we only report the results of the baseline model without reinforcement learning. CL4AC~\cite{LiuHMKTPW21} is based on contrasting learning to reduce the domain difference by learning the correspondence between the audio clips and captions. AT-CNN10~\cite{xu2021investigating} method mainly uses transfer learning to initialize the encoder parameters of the audio captioning by learning the local feature from the audio tagging task and the global feature from acoustic scene classification. 
In addition, we consider a Pre-Bert method for audio captioning as a baseline where the Pre-trained BERT language is used as the decoder for the AudioCaps dataset \cite{Liuprebert22}. Table~\ref{tab:audioresult} shows the performance comparison of the proposed and baseline models. It can be seen that the proposed LHDFF outperforms the baseline models. Moreover, Table~\ref{tab:audio1} shows the proposed LHDFF can learn more audio information than the baseline model only uses the high-dimensional feature. 

\textbf{High- vs. fusion and low- embedding feature} We report the result of the proposed LHDFF model using only the fusion embedding feature as the input to a single transformer decoder, i.e., the LHDFF (fusion) in Table \ref{tab:audioresult}. In addition, in Table \ref{tab:audioresult}, we report the result of the Baseline that only outputs a high-dimensional feature in the encoder, which is then input to the transformer decoder. It can be observed that the performance of the baseline model with only a high dimensional feature is worse than the LHDFF model which uses only the fusion embedding feature. Moreover, we also report the result of the LHDFF using the high- and low-dimensional features as the inputs to the dual transformer decoder, i.e., LHDFF~(nonfusion). From the results, it is easy to find that the performance of LHDFF~(nonfusion) is worse than the Baseline and LHDFF fusion methods. These results show that the low-dimensional feature can learn limited useful information. But, when we use the fusion way that can allow the low- and high-dimensional features to cooperate and complement each other that can better emphasize some audio scenes.

\textbf{Performance analysis of different low-dimensional features} While the encoder of the proposed LHDFF consists of four convolutional layers, the low-dimensional feature is obtained from the third convolutional block. To test the efficiency of low-dimensional features from different convolutional blocks, we report the result of the proposed LHDFF using the low-dimensional feature from the second convolutional block, i.e., LHDFF~(fusion block2). Therefore, in the encoder of LHDFF~(fusion block2), the fused feature of LHDFF~(fusion block2) is fused by the low-dimensional feature from the second convolutional block and the high-dimensional feature. From Table~\ref{tab:audioresult}, we can find that the proposed LHDFF outperforms the LHDFF~(fusion block2) which implies that the low-dimensional feature from the third convolutional block includes more useful information than the low-dimensional feature from the second convolutional block. 

\textbf{Dual decoder vs. Single decoder} We also evaluate the effectiveness of the dual decoder in the proposed LHDFF, as compared with those of the Baseline and the LHDFF~(fusion) that use a single decoder using the high-dimensional feature and fusion embedding feature. From the results, we can find that the proposed LHDFF with dual decoder outperforms the Baseline and LHDFF~(fusion) which are based only on a single decoder. The main motivation behind using a dual decoder is similar to ensemble learning which aims to utilize multiple algorithms to achieve better performance than can be obtained with a single algorithm. 

\section{Conclusion}
\label{sec:conc}

A new encoder-decoder model called the LHDFF model is proposed for AAC. In LHDFF, a residual PANNs encoder (RPANNs) is proposed by fusing the low-dimensional feature of the third convolutional block and the high-dimensional feature from the final layer of PANNs. The low-dimensional approach successfully overcomes the high-dimensional problem of learning the audio scenarios with available training data by learning models, it only works with specific audio scene types. In addition, there is always insufficient information learning of high-dimensional methods owing to high-dimensional representations having a large amount of information. The fusion of the low- and high-dimensional features allows the low- and high-dimensional features to cooperate and complement each other, which can emphasize the reoccur audio scenes. In addition, a dual transformer decoder is designed to generate the captions from these features in parallel. In a dual transformer decoder, a probabilistic approach is designed to fuse the outputs of the two transformer decoders. Experimental results show that LHDFF achieves considerable improvements over the existing audio captioning models.
 
\section{Acknowledgement}
This work is partly supported by UK Engineering and Physical Sciences Research Council (EPSRC) Grant EP/T019751/1 ``AI for Sound'', a Newton Institutional Links Award from the British Council, titled ``Automated Captioning of Image and Audio for Visually and Hearing Impaired" (Grant number 623805725), British Broadcasting Corporation Research and Development~(BBC R\&D), a PhD scholarship from the University of Surrey, and a Research Scholarship from the China Scholarship Council (CSC). For the purpose of open access, the authors have applied a Creative Commons Attribution (CC BY) licence to any Author Accepted Manuscript version arising.

\bibliographystyle{IEEEtran}
\bibliography{mybib}
\balance

\end{document}